\documentclass[10pt,epj,tightenlines,twocolumn,eqsecnum,superscriptaddress,floats,showpacs,aps,nofootinbib,altaffilletter,secnumarabic]
{svjour}
\usepackage{graphicx}
\usepackage{dcolumn}
\usepackage{bm}
\usepackage{amssymb,amsmath,latexsym,amsfonts}
\usepackage{color}
\usepackage{dcolumn}
\usepackage{epsf}
\usepackage{float}
\usepackage{hyperref}
\usepackage{flushend}
\usepackage{cuted}

\makeatletter\@addtoreset{equation}{section}

\begin{document}
\title{Stability Analysis of a Bose--Einstein Condensate Trapped in a Generic Potential}
\author{El\'ias Castellanos\inst{1}, Celia Escamilla--Rivera\inst{1} \and  Mayra J. Reyes--Ibarra\inst{2}%
%
}                     
%
%
\institute{Mesoamerican Centre for Theoretical Physics, Universidad Aut\'onoma de Chiapas.
Ciudad Universitaria, Carretera Zapata Km. 4, Real del Bosque (Ter\'an), 29040, Tuxtla Guti\'errez, Chiapas, M\'exico. \and Departamento de
F\'{\i}sica, Centro de Investigaci\'on y Estudios Avanzados del IPN. 
 A. P. 14--740,  07000, Ciudad de M\'exico, M\'exico.}
\date{Received: date / Revised version: date}
%
\abstract{
We investigate the dynamical behavior of the Gross--Pitaevskii equation for a 
Bose--Einstein condensate trapped in a spherical power law potential restricted to the repulsive case, from the \emph{dynamical system formalism} point of view. A five--dimensional dynamical system is found (due the symmetry of the Gross--Pitaevskii equation interacting with a potential), where the Thomas--Fermi approximation constrains the parameter space of solutions. We show that for values of the power law exponent equal or smaller than 2 the system seems to be stable. However, when the corresponding  exponent is bigger than 2, the instability of the system grows when the power law exponent grows, indicating that large values of the aforementioned parameter can be related to a loss in the number of particles from the condensed state. This fact can be used also to show that the stability conditions of the condensate are highly sensitive to the exponent associated with the external potential.
} 
\authorrunning{El\'ias Castellanos et. al.}
\titlerunning{Stability Analysis of a Bose--Einstein Condensate Trapped in a Generic Potential}
\maketitle

\section{Introduction}

The phenomenon of Bose--Einstein condensation \cite{Bose,Bose1,Bose2} is one of the most remarkable N--body quantum 
phenomena that can be observed at macroscopic scales. Since its laboratory observation Bose-Einstein 
condensation of dilute atomic gases has stimulated an enormous amount of related work  \cite{Anderson,A3,A4,A5} . We found mathematical issues related to Bose--Einstein condensates \cite{Lieb}, several theoretical and heuristic aspects \cite{Dalfovo,S}, 
also as test tools in gravitational physics, or even to model dark 
matter in the universe. The study of its associated thermodynamic properties is 
naturally also a very pertinent aspect \cite{Dalfovo,Pethick,ueda,ref,ref1,ref2}. The experimental 
understanding of Bose--Einstein condensates can be achieved in several alkali atoms which allows 
the use of various atomic systems, for instance, $^{87}$Rb, $^{23}$Na, which have repulsive interactions within the condensate \cite{Pethick,TA}. Conversely, $^{7}$ Li and $^{85}$Rb that have attractive interactions \cite{Pethick,TA}. The low energy properties and specifically the stability of the condensate properties can be characterized through the s--wave 
scattering length. These conditions of the condensate trapped in magnetic traps has been extensively analyzed from the numerical and analytical point of view \cite{TA}. It is seems that the condensate was predicted to be (meta)stable in magnetic traps, when the number of atoms is below some critical number \cite{K}. In other words, the possible values of the s--wave scattering length are intimately linked to the stability of the system under these circumstances. However, there is at least one more parameter linked to the stability of the condensate, the specific form of the trapping potential. The form of the trapping potential is strongly related to the macroscopic behavior of the condensate, and clearly the properties of the \emph{condensed cloud} must be highly sensitive to its shape \cite{dm}. Then, it is interesting to analyze the stability of the condensate when one varies the shape of the trap besides the value of the scattering length. In this work, we will assume that the value of the scattering length is positive, i.e., we assume from the very beginning that our system is a repulsively condensate. However, there are in the literature several works related to the stability analysis when the value of the scattering length is negative \footnote{For works related to the stability of  attractively interacting condensates, see refs.\,\cite{ATRA,ATRA1,ATRA2,ATRA3,ATRA4}}. We must also mention that experimental realizations of Bose Einstein condensates in higher power law potential traps has been extensively investigated, see for instance refs.\,\cite{EXP,EXP1,EXP2,EXP3}. 

Additionally, the dynamical system formalism \cite{libro0} is a mathematical tool that is implemented in order to rewrite the evolution equations as plane-autonomous system and to analyze the stability of the equation under consideration. The results of the latter can give some information about the physics of the system behaviour which can be hidded in the corresponding critical points of the system.

Therefore, we will compare the stability conditions of the condensate with the stability of the Gross--Pitaevskii equation
for a Bose--Einstein condensate trapped in a spherical power law potential. As far as we know, this mathematical tool has never been used in this context. Thus, the analysis of stability conditions of the Gross--Pitaevskii equation as an autono\-mous system over the condensate in a generic potential can be helpful in order to find the most favorable scenarios in which the system actually acquires more stable configurations, when repulsive interactions are present.   

The outline of the paper is as follows: in Section \ref{sec:prop} we describe the main properties of the Gross--Pitaevskii equation and of the corresponding spherical power law potential. Also, some issues regarded the Thomas--Fermi approximation are revised.
Section \ref{sec:dynamical} is dedicated to the dynamical analysis of the cases in which a Bose--Einstein condensate is endowed in a power law potential. We first show that there are four cases of interest, related to the values of the corresponding exponent $s$. To have a complete picture of the solution for the Gross--Pitaevskii equation in a Thomas--Fermi approximation,
these equations are written as an autonomous system, and we study its critical points and general
trajectories in the phase space of the resulting dynamical variables. 
In Section \ref{sec:gen_pot} we analyze some thermodynamical properties of the system and its relation with the results obtained in dynamical analysis. 
Finally, Section \ref{sec:Conclusions} is devoted to final comments about the general properties of the Bose--Einstein condensate trapped in a spherical power law potential. 


\section{The spherical power law potential and the Gross--Pitaevskii equation}
\label{sec:prop}
Let us start with the time dependent Gross--Pitaevskii equation in 3--dimensions

\begin{eqnarray}
\label{eq:GP}
i\hbar \partial_{t}\psi\, (r,t) &=& - \frac{\hbar^2}{2m} \nabla^2\, \psi(r,t) + V(r)\, \psi(r,t) 
\nonumber \\ 
&&+ U_0\, |\psi(r,t)|^2  \, \psi (r,t) ,  
\end{eqnarray}
where $m$ is the atomic mass and $U_0= 4\pi \hbar^2 a_s/m$ describes the interaction between the particles in the condensate with $a_s$  the $s$--wave scattering length, which in this approach is assumed to be positive. The validity of the Gross--Pitaevskii  equation is based on the condition that the $s$-wave scattering length be much smaller than the average distance between atoms and that the number of atoms in the condensate be much larger than one. The Gross--Pitaevskii  equation can be used, at temperatures below the condensation temperature, to explore the macroscopic behavior of the system, characterized by variations of the order parameter $\psi(r,t)$ over distances larger than the mean distance between atoms.

The trapping potential in Eq.\,(\ref{eq:GP}) is given by:
\begin{equation}\label{eq:Tpotential}
V(r)=A\left(\frac{r}{a_{0}}\right)^{s}.
\end{equation}
The potential (\ref{eq:Tpotential}) is the so--called spherical power law potential where, for simplicity, we assume spherical symmetry (see for instance Refs. \cite{zobay,zobay1,zobay2} and references therein for more general potentials). 
In the generic potential (\ref{eq:Tpotential}) we have $A=\hbar\omega_{0}/2$ and $a_{0}=\sqrt{\hbar/m \omega_{0}}$, which are characteristic scales of energy and length associated with the trap \cite{eli,eli1}, with $\omega_{0}$ the corresponding frequency and $\hbar$ the reduced Planck's constant. The parameter $s$ depicts the exponent of the spherical power law potential, which in our approach will play a relevant role in the dynamics. Notice that if we set $s=2$ in the spherical power law potential (\ref{eq:Tpotential}), we recover the more common harmonic trap. These type of potentials are known as generic spherical potentials. It is quite interesting to analyze the behavior of the condensate when the shape of the trap varies as a power of the radial coordinate for several reasons. This class of trapping potentials could be useful in order to cool the system with an adiabatic procedure, by changing the shape of the trap \cite{Dalfovo}. The relevant thermodynamical properties associated with the system, e.g., the condensation temperature, the number of particles within the condensate, the corresponding density, the entropy, etc., explicitly exhibits a sensitive trap--dependence, i.e., the macroscopic behavior of the condensate is highly sensitive to the shape of the external trapping potential. In this sense, it is relevant to analyze the stability of the system when this spherical power law potential is present. 

In order to obtain a dimensionless version of the Gross--Pitaevskii equation (\ref{eq:GP}), let us define, as usual, the following dimensionless variables in function of the characteristic scales associated with the system  \cite{M.M,M.M1,M.M2}
\begin{eqnarray}
\bar{t} &=& \omega \,  t \, , \quad \bar{r} = \frac{r}{a_0} \, , \nonumber\\ 
 \bar{\psi}(\bar{r},\bar{t}) &=& a^{3/2}_0 \psi (r,t) \, ,  \quad \bar{V}(\bar{r}) = \frac{V (r)}{2A} \, .
\label{eq:variables}
\end{eqnarray}
By using the dimensionless variables defined above, the Gross--Pitaevskii equation in 3--dimensions becomes
\begin{eqnarray}
i \partial_{\bar{t}}\bar{\psi}\, (\bar{r},\bar{t})&=& - \frac{1}{2} \bar{\nabla}^2\, \bar{\psi}(\bar{r},\bar{t}) + \bar{V}(\bar{r})\, \bar{\psi}(\bar{r},\bar{t}) 
\nonumber\\ &&+ \beta \, |\bar{\psi}(\bar{r},\bar{t})|^2 \, \bar{\psi} (\bar{r},\bar{t}) \, ,
\label{eq:dimensionless}
\end{eqnarray}
where we have defined $\beta=U_{0}/ \hbar \omega_{0} a^{3}_{0}$. As was mentioned above, we restrict our analysis to positive values of the corresponding self--interaction potential since positive scattering lengths provides repulsive interactions among the particles. 

Consequently,  the time independent dimensionless Gross--Pitaevskii equation can be obtained by written the dimensionless order parameter $\bar{\psi}(\bar{r},\bar{t})$ as follows:
\begin{eqnarray}
\label{eq:stationary}
\bar{\psi}(\bar{r},\bar{t})=e^{-i\bar{\mu}\bar{t}} \,\bar{\phi}(\bar{r}) \, ,
\end{eqnarray}
where we have defined the dimensionless chemical potential as $\bar{\mu}=\mu/\hbar \omega_{0}$. Inserting (\ref{eq:stationary}) into (\ref{eq:dimensionless}), this leads to the following equation

\begin{eqnarray}\label{eq:GP2}
 \bar{\nabla}^2\, \bar{\phi}(\bar{r})
= 2\left[\bar{V}(\bar{r})-\bar{\mu}\right]\bar{\phi}(\bar{r}) +2\beta\bar{\phi}^3 (\bar{r}).
\end{eqnarray}
where the Laplacian operator is given by $\bar{\nabla}^2=\bar{r}^{-2}\partial_{\bar{r}}\bar{r}^{2}\partial_{\bar{r}}$. Consequently Eq.\,(\ref{eq:GP2}) reads
\begin{eqnarray}
\label{eq:GP3}
 \frac{d^{2}\bar{\phi}(\bar{r}) }{d\,\bar{r}^{2}}+\frac{2}{\bar{r}}\frac{d\,\bar{\phi}(\bar{r}) }{d\,\bar{r}}
= 2\left[\bar{V}(\bar{r})-\bar{\mu}\right]\bar{\phi}(\bar{r}) +2\beta\bar{\phi}^3 (\bar{r}).
\end{eqnarray}

An analytical and formal solution associated with our system can be obtained by neglecting the kinetic energy term from the very beginning in Eq.\,(\ref{eq:GP2}), this is the so--called Thomas--Fermi approximation \cite{Pethick}, with the result
\begin{equation} \label{eq:TFA}
n(\bar{r})=|\bar{\phi}(\bar{r})|^{2}=\frac{\bar{\mu}-\bar{V}(\bar{r})}{\beta},
\end{equation}
while $n(r)=0$ outside of this region \footnote{
We are able to estimate the spatial extend $\bar{r}_{e}$ of the cloud in the dimensionless variables, through the condition $\bar{\mu}=\bar{V}(\bar{r})$, which in our case is given by
\begin{equation}
\bar{r}_{e}=(2\bar{\mu})^{1/s}.
\end{equation}  
This spatial extend fixes the regime of validity of the Thomas--Fermi approximation and consequently, also fixes the validity of the analysis followed in this paper.}. Here, $n(\bar{r})=|\bar{\phi}(\bar{r})|^{2}$ depicts the density of particles within the condensate in the Thomas--Fermi approximation. Therefore, the Thomas--Fermi approach is valid for systems at very low temperatures $T<T_{c}$, where $T_{c}$ is the condensation temperature, for weakly interacting systems, for sufficient large clouds, and when the kinetic energy is negligible with respect to the potential and the interaction ones. Additionally, the Thomas--Fermi approximation is valid near to the minima of the potential. Finally, we must add that the Thomas-Fermi approximation fails for trapped condensates near the edge of the cloud, due to the divergent behavior of the kinetic energy (i.e. the total kinetic energy per unit area diverges on the boundary of the system) \cite{DAL1}.

\section{Dynamical system structure}
\label{sec:dynamical}

The first step to approach to our goal consists in study the evolution of the Gross--Pitaevskii equation (\ref{eq:GP2}) in which the condensate is endowed with the spherical power law potential (\ref{eq:Tpotential}). As in the standard case, it is possible to perform a dynamical study
of the model so that its relevant solutions are easily unveiled \cite{Copeland:1997et,EscamillaRivera:2010py,Boehmer:2014vea}. We will see that the Thomas--Fermi approximation leads to a constraint equation on the parameter space whose solution defines a subspace where stable solutions of the system are possible. In order to construct a dynamical system for a Bose--Einstein condensate trapped in the spherical power law potential  (\ref{eq:Tpotential}), we first introduce a set of conveniently chosen variables which may allow rewriting the evolution equation as an
autonomous system subject to the constraint arising from the Thomas--Fermi approximation.
We choose the following variables
\begin{equation}\label{eq:dynamical}
x\equiv \bar{\phi}^\prime, \quad y\equiv \frac{\bar{r}^s}{2} -\bar{\mu}, \quad z\equiv \bar{\phi}, \quad w\equiv\frac{s}{2}\left(\frac{\bar{r}^s}{\bar{r}}\right), \\
v= \frac{2}{\bar{r}}\bar{\phi},
\end{equation}

where the prime denote derivatives with respect to $\bar{r}$. Let us clarify that these \textit{dynamical} coordinates correspond to the phase space.
In these variables  the Thomas--Fermi approximation constraint is implemented by setting
\begin{equation}\label{eq:TFA_constraint}
F(y,z):=\kappa |z|^2 = -2y,
\end{equation}
where $\kappa\equiv 2\beta$.
We shall restrict ourselves to the parameter space delimited by this hyperbolic region, which contains the solutions in this approximation
for the Gross--Pitaevskii equation. Combining (\ref{eq:GP3}) and the variables (\ref{eq:dynamical}), the equations of motion read
\begin{subequations}
\begin{eqnarray}
x^{\prime} &=& 2yz +\kappa z^3, \\ \label{eq:systemdynamical-a}
y^{\prime} &=& w, \\ \label{eq:systemdynamical-b}
z^{\prime} &=& x, \\ \label{eq:systemdynamical-c}
w^{\prime}&=& \frac{s(s-1)}{\bar{r}^2}\left(y+\bar{\mu}\right), \label{eq:systemdynamical-d} \\
v^{\prime}&=& \frac{2}{\bar{r}}(x-z).
\end{eqnarray}\label{eq:systemdynamical}
\end{subequations}
The stability of the Gross--Pitaevskii equation through its critical points is investigated by using linear perturbations (See Appendix \ref{app1}). 
The results of the analysis of the dynamical system (\ref{eq:systemdynamical}), their critical points and its stability properties, are summarized in Table \ref{tab:critical}. At this point we are interested in where the trapping potential, (i.e., for different values of the exponent $s$), can set a stable (or unstable) scenario. To proceed with these ideas, we follow the cases related to the Bose--Einstein condensate physics. Notice that if we set $s=0$ in the spherical power law  potential (\ref{eq:Tpotential}), we obtain a free gas. For $s=1$ a linear trap and for $s=2$ the usual harmonic oscillator. Additionally, when $s\rightarrow \infty$, we obtain the box potential. 

Let us now establish the stability of the Gross--Pitaevskii equation conditions for some potentials of interest. 

\begin{itemize}
\item Case $s=0$, i.e $\bar{V}=\text{const}$. The autonomous system is reduced to the following three dimensional system:
\begin{eqnarray}
x^{\prime} = 2\alpha_1 z+\kappa z^3, \quad
z^{\prime} =x, \quad
v^{\prime}&=& \frac{2}{\bar{r}}(x-z),
\end{eqnarray}
where $\alpha_1 =1/2 -\bar{\mu}$. The real critical point for this system is $(x_c =0,z_c=0, v_c=0)$. The constraint is fixed by $|z|^2 = -2\alpha_1 /\kappa$. 
\begin{figure*}
\centering
\includegraphics[width=7.1cm]{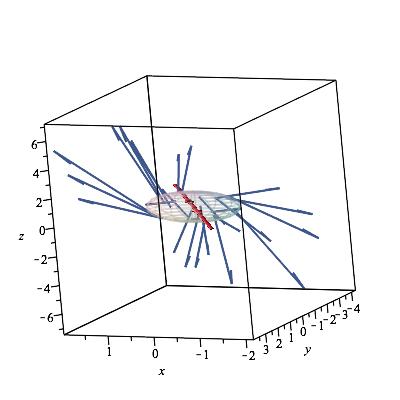}
\centering
\caption{\label{fig:case_szero}\label{fig:case_sone} Phase portrait for case $s=1$. The flux of the dynamical system (represented by the blue arrows) in represented by the cases $\xi_{+}$ with vector fields for $m_{+}$ for the case $s=1$ given in Table 1. We notice that now the autonomous system is in  3--dimensions due the quadratic term in the potential. We infer from this field plot the region where the stability of the Gross--Pitaevskii equation can be achieved. The ellipsoidal structure (indicated by the ellipsoidal region in smooth colors) is delimited by the Thomas-Fermi constrain $|z|^2 = (2\bar{\mu}-\bar{r}^2)/\kappa$. The axes represent the dynamical variables in 3-dimensiones with $s=1$.}
\end{figure*}
The trajectories in Figure \ref{fig:case_szero} for the case 
$\xi_{+}$ of $m_{+}$ 
start at the critical point then diverge to infinite-distance out. As a result, the critical point for the case results in a saddle point, which is always unstable. When $s=0$ the potential becomes in a constant that can be re-absorbed in the dimensionless chemical potential. This type of potential may not be of physical relevance. However, we have included this case for mathematical reasons. In this scenario, the system is unstable, according to this approach, which makes sense since there is no confinement region. This behavior is showed in Figure \ref{fig:case_sone}.

\item Case $s=1$, i.e $\bar{V}=\bar{r}/2$, which corresponds to the linear trapping potential. The autonomous system is reduced to the following set of equations:
\begin{eqnarray}
&&x^{\prime} = 2yz +\kappa z^3 = 2\Big(\frac{\bar{r}}{2}-\bar{\mu}\Big)z+\kappa z^3, \nonumber \\ && y^{\prime}=1/2, \quad z^{\prime}=x,
\quad v^{\prime}= \frac{2}{\bar{r}}(x-z),
\end{eqnarray}
with two critical points at $(x_{c_{1}} =0, y_{c_1}=y, z_{c_{1}}=0, v_{c_{1}}=0)$ and $(x_{c_2} =0, y_{c_2}=-\kappa z^2/2, z_{c_2}=z, v_{c_2}=2z_{c_{2}}/\bar{r})$ The constraint is fixed by the curve $|z|^2 = (2\bar{\mu}-\bar{r})/\kappa$. 
The instability happens only along the eigenvectors corresponding to $\pm\sqrt{2y}$ and $\pm\sqrt{2\kappa}z$, which are unstable (saddle) points. The first eigenvalues are null, and then the system is indifferent under perturbations along the constraint curve for this case.
Also we notice that for a vanishing potential $\bar{\mu}\rightarrow 0$, the fourth dynamical variable is $v=z/y$, therefore fixing a new \textit{minimal} constraint in one dimension according to (\ref{eq:TFA_constraint}) as $F(z):=4\beta z^3 =- v_{(s=1,\bar{\mu}\rightarrow 0)}$.
The phase space of solutions for this case is reduced to a one-dimension in comparison with Eq.(\ref{eq:TFA_constraint}). Inside this constriction, only the unstable solutions with eigenvalues 
$\pm\sqrt{2\kappa}z$ are allowed. Additionally, according to the spatial extend for this case $\bar{r}_{e}=2\bar{\mu}$, the above constraint indicates that when $\bar{\mu}\rightarrow 0 $ then $\bar{r}_{e}$ decreases with respect to the $\bar{r}_{e}$ obtained from Eq.\,(\ref{eq:TFA_constraint}).

\item Case $s=2$, i.e $\bar{V}=\bar{r}^2/2$, which corresponds to the more common harmonic oscillator potential. The autonomous system is reduced in this case to the system:
\begin{eqnarray}
&&x^{\prime} = 2yz +\kappa z^3 = 2\Big(\frac{\bar{r}^2}{2}-\bar{\mu}\Big)z+\kappa z^3, \nonumber\\ && y^{\prime}=w, \quad z^{\prime}=x, \quad w^{\prime}=1, \quad v^{\prime}= \frac{2}{\bar{r}}(x-z)
\end{eqnarray}
with two critical points at $(x_{c_{1}} =0, y_{c_1}=y, z_{c_{1}}=0, w_{c_1}=0, v_{c_1}=0)$ and $(x_{c_{2}} =0, y_{c_{2}}=-(\kappa z^2)/2, z_{c_{2}}=z, w_{c_2}=0, v_{c_2}=2z_{c}/\bar{r})$. The constraint is fixed by the curve $|z|^2 = (2\bar{\mu}-\bar{r}^2)/\kappa$. For the origin critical point we observe that only the non-vanishes eigenvalues, which points in $z$-direction and $v$-direction gives the information about the stability of this point. In such case, when $\bar{r}>2\bar{\mu}$ ($\bar{r}<2\bar{\mu}$) the point will be unstable (stable).  For the case $s=2$, or the usual harmonic trap, the system seems to be stable in all cases. In other words, note that in our approach the Thomas--Fermi approximation fixes the permisible values for the spatial extend of the cloud as $\bar{r}_{e}=(2\bar{\mu})^{1/s}$, this means that  for all $\bar{r}$ of interest $\bar{r}<(2\bar{\mu})^{1/s}$, and consequently $\bar{r}<2\bar{\mu}$ is always satisfied, leading to a stable scenario.     

\item Case $s=3$, i.e $\bar{V}=\bar{r}^3/2$, which corresponds to a cubic potential. The autonomous system is reduced to the system:
\begin{eqnarray}
&& x^{\prime} = 2yz +\kappa z^3 = 2\Big(\frac{\bar{r}^3}{2}-\bar{\mu}\Big)z+\kappa z^3, \nonumber\\ && y^{\prime}=w= \frac{3}{2}\bar{r}^2, \quad z^{\prime}=x, \quad w^{\prime}=\frac{6}{\bar{r}^2}(y+\bar{\mu})\nonumber \\
&& v^{\prime}= v+\frac{2}{\bar{r}}z'.
\end{eqnarray}
with three critical points at $(x_{c_{1}} =0, y_{c_1}=-\bar{\mu}, z_{c_{1}}=0, w_{c_1}=0,v_{c_1}=0)$ and $(x_{c_{(2,3)}} =0, y_{c_{(2,3)}}=-\bar{\mu}, z_{c_{(2,3)}}=\pm\sqrt{\frac{2\bar{\mu}}{\kappa}}, w_{c_{(2,3)}}=0, v_{(2,3)}=0)$. The constraint is fixed by the curve $|z|^2 =2\bar{\mu}/\kappa- 8z^3/v^3$. To have a Bose--Einstein condensate the values of $\bar{\mu}>0$, i.e the first energy state is a non-vanishing constant. Then, for the first critical point the trajectories stay in an elliptical traces. However, with each revolution, their distance from $c_1$ grow/decay exponentially according to the term $e^{m_{+}\bar{r}}$ (see (\ref{eq:solutions_dynamical})). Therefore, the 5-dimensional phase portrait shows trajectories that attractor
to the critical point 
 (when $m_{c_1}>0$). Or trajectories that loiter to the attractor converge to the critical point (when $m_{c_1}<0$). As a result, when $\bar{\mu}>0$, $c_1$ is a attractor point, which it is asymptotically stable if $m_{c_1}<0$ and it is unstable if $m_{c_1}>0$.

\item Case $s=4$, i.e $\bar{V}=\bar{r}^4/2$, which corresponds to the quartic potential. The autonomous system is now reduced to the following system of equations:
\begin{eqnarray}
&&x^{\prime} = 2yz +\kappa z^3 = \Big(\bar{r}^4-2\bar{\mu}\Big)z+\kappa z^3, \nonumber\\&& y^{\prime}=w= 2\bar{r}^3, \quad z^{\prime}=x, 
\nonumber\\ && w^{\prime}=\frac{12}{\bar{r}^2}(y+\bar{\mu})=6\bar{r}^2,  \quad v^{\prime}= v+\frac{2}{\bar{r}}z'
\end{eqnarray}
with four critical points at $(x_{c_{(1,2)}} =0, y_{c_{(1,2)}}=\pm\bar{\mu}, \\z_{c_{(1,2)}}=0, w_{c_(1,2)}=0)$ and $(x_{c_{(3,4)}} =0, y_{c_{(3,4)}}=-\bar{\mu}, z_{c_{(3,4)}}\\=\pm\sqrt{\frac{2\bar{\mu}}{\kappa}}, w_{(3,4)}=0, v_{(3,4)}=0)$. The constraint is fixed by the curve 
$|z|^2 =2\bar{\mu}/\kappa- 8z^4/v^4$. As in the case $s=3$, for the first critical point the trajectories are elliptical. Therefore, we have the same 5-dimensional phase portrait with spiral trajectories when $\bar{\mu}>0$, $c_1$ is a spiral point, which it is asymptotically stable if $m_{c_1}<0$ and it is unstable if $m_{c_1}>0$. Notice that $m_{c_1}$ is always positive according to Table\,(\ref{tab:critical1}) and then is unstable. In comparison the latter case, we have a critical point for $\bar{\mu}<0$, which means $\bar{r}_e =-2\bar{\mu}^{1/4}$, which is a mathematical solution but is unphysical since $\bar{\mu}$ is always positive for temperatures smaller that the condensation temperature $T_{c}$.
\end{itemize}
Furthermore, for the cubic potential $s=3$ and the quartic potential $s=4$ the numerical stability analysis shows that these systems present an instability due to the presence of 
attractors trajectories. If we interpreted these trajectories as the velocity field of the particles in the condensate, we can immediately observe that when $s$ grows the instability increases.
These facts can be interpreted as follows, when $s>2$ grows, the particles forming the condensate tends to leave the ground state to form a sea of excited particles, which interacts with the condensate affecting its stability. 

Finally, let us remark that due to the corresponding limitations associated with the Thomas--Fermi approach used in the above analysis, the deduced instabilities are valid in a certain region near to the minima of the potential, i.e., for the region $\bar{r}<(2\bar{\mu})^{1/s}$. Thus, as a consequence we can not extrapolate our results to the entire condensate.

\begin{table}
 \centering
 \caption{Critical points for the autonomous system (\ref{eq:systemdynamical}) in where exist solutions related to the stability of the Gross--Pitaevskii equation}
 \medskip
 \begin{tabular}{@{}ccc@{}}
 \hline
Label &Critical point\\
&($x_{c}, y_{c}, z_{c}, w_{c}, v_{c} $) \\
 \hline
  &     \\
A&($0,-\bar{\mu}^2,0,0,0$)   \\
B&($0,-\bar{\mu}^2, \pm \sqrt{\frac{2}{\kappa}}\bar{\mu}, 0, 0$)  \\
  &     \\
\hline
\end{tabular}
\label{tab:critical}
\end{table}
\begin{table}
 \centering
 \caption{The corresponding eigenvalues related to the critical points.}
 \medskip
 \begin{tabular}{@{}ccc@{}}
 \hline
Label &Eigenvalues\\
&($m_1, m_2, 
m_3, m_4, m_5$) \\
 \hline
  &     \\
A  & $\left(\frac{\sqrt{s^2 -s}}{\bar{r}},-\frac{\sqrt{s^2 -s}}{\bar{r}},\sqrt{2\bar{\mu}}i, \sqrt{2}\bar{\mu} i, 2\sqrt{2}\bar{\mu} i/\bar{r}\right)$  \\
B& $\left(\frac{\sqrt{s^2 -s}}{\bar{r}},-\frac{\sqrt{s^2 -s}}{\bar{r}},\sqrt{2\bar{\mu}}, \sqrt{2}\bar{\mu},  2\sqrt{2}\bar{\mu}/\bar{r}\right)$  \\
  &     \\
\hline
\end{tabular}
\label{tab:critical1}

\end{table}

\section{Condensation temperature and condensate fraction}
\label{sec:gen_pot}

The second step to approach to our goal consists of an analysis of some of the thermodynamic properties of our bosonic gas trapped in the spherical power law potential Eq.\,(\ref{eq:Tpotential}) in order to study its relation to the dynamical structure analyzed in the previous section. The properties of a condensate trapped in a spherical power law potential have been extensively analyzed, see for instance, \cite{zobay,eli,H,L,sal} and references therein.

To this aim, let us propose a particularly simple Har\-tree--Fock type
spectrum, in the semi--classical approximation. This approach basically consists of the
assumption that the constituents of the gas behave like non--interacting bosons moving in a self--consistent mean field, valid when the semiclassical energy spectrum $\epsilon_{p}$ is bigger than the corresponding chemical potential $\mu$, for dilute gases \cite{Dalfovo,Pethick}
\begin{equation}
\label{HF0} \epsilon_{p}=\frac{p^{2}}{2m}+ V(r)+2U_{0}n(r),
\end{equation}
where $p$ is the momentum and $m$ is the mass of a single particle respectively. The term $2U_{0}n(\vec{r})$ is a mean field generated by the interactions with the other constituents of the bosonic gas, being $n(r)$ the spatial density of the cloud \cite{Dalfovo}.

In the semiclassical approximation, the single--particle
phase--space distribution may be written as \cite{Dalfovo,Pethick}
\begin{equation}
\label{SPSD}
n(r,p)=\frac{1}{e^{\beta(\epsilon_{p}-\mu)}-1},
\end{equation}
where $\beta=1/\kappa T$, $\kappa$ is the Boltzmann constant, $T$ is the temperature, and $\mu$ is the chemical potential.
The number of particles in the 3--dimensional space obeys the
normalization condition \cite{Dalfovo,Pethick},
\begin{equation}
 N=\frac{1}{(2 \pi \hbar)^{3} }\int d^{3}r\hspace{0.1cm} d^{3} p\hspace{0.1cm} n(r,p),
\label{NC}
\end{equation}
where
\begin{equation}
\label{n1} 
n(r)=\int  d^{3} p \hspace{0.1cm}
n(r,p),
\end{equation}
is the corresponding spatial density. Using expression
(\ref{HF0}), and integrating expression (\ref{SPSD}) over momentum, with the help of (\ref{n1}), we get the spatial
distribution associated with our semi--classical spectrum
(\ref{HF0})
\begin{equation}
\label{MSD10}
 n(r)=\lambda^{-3} g_{3/2}\Bigl(e^{\beta[\mu -V(r)-2U_{0}n(r)]}\Bigr)
\end{equation}
where $\lambda=\Bigl(\frac{2 \pi \hbar^{2}}{m \kappa
T}\Bigr)^{1/2}$, is the de Broglie thermal wavelength. The function $g_{\nu}(z)$ is the so--called Bose--Einstein function defined
by \cite{Phatria}
\begin{equation}
g_{\nu}(z)=\frac{1}{\Gamma(\nu)}\int_{0}^{\infty}\frac{x^{\nu-
1}dx}{z^{-1}e^{x}-1}. \label{BEF}
\end{equation}
By using the properties of the Bose--Einstein functions \cite{Phatria}, we are able to expand expression (\ref{MSD10}) to first order in $U_{0}$, with the result
\begin{eqnarray}
\label{MSD1} n(r)&=& n_{0}(r)\Bigg(1-\frac{2U_{0}}{\kappa T} \Bigl(\frac{m\kappa T}{2\pi \hbar^{2}}\Bigr)^{3/2}g_{1/2}(Z)\Bigg),
\end{eqnarray}
where
\begin{equation}
 Z=e^{\frac{1}{\kappa T}[\mu-V(r)]},
\end{equation}
being $n_{0}(r)$ the space density distribution for the ideal
case $U_{0}=0$,
\begin{equation}
 \label{n0}
n_{0}(r)=\lambda^{-3} g_{3/2}(Z).
\end{equation}

Integrating the normalization condition (\ref{NC}) and using expression
(\ref{MSD1}) with the corresponding potential (\ref{eq:Tpotential}), this allows
us to obtain an expression for the number of particles as a function
of the chemical potential $\mu$, the temperature $T$, the coupling
constant $U_{0}$
\begin{eqnarray}
\label{NPINT} N &=& N_{0}+\Bigl(\frac{\kappa T}{\hbar \omega_{0}}\Bigr)^{\frac{3}{2}+\frac{3}{s}} \frac{2^{\frac{3}{s}} \Gamma(\frac{3}{s}+1)}{2^{\frac{3}{2}} \Gamma(\frac{3}{2}+1)} g_{\frac{3}{2}+\frac{3}{s}}(z) \nonumber\\ &\times& \Bigg[1-2U_{0}(\kappa T)^{\frac{1}{2}}\Bigl(\frac{m}{2\pi\hbar^{2}}\Bigr)^{\frac{3}{2}}\,\frac{G_{3/2,1/2,3/s}(z)}{g_{\frac{3}{2}+\frac{3}{s}}(z)}\Bigg],
\end{eqnarray}
with the following definition
\begin{equation}
G_{3/2,1/2,3/s}(z)= \sum_{i=1}^{\infty} \sum_{j=1}^{\infty} \frac{z^{(i+j)}}{i^{3/2}j^{1/2}(i+j)^{3/s}},
\end{equation}
where $\Gamma(x)$ is the Euler function, $N_{0}$ the number of particles in the grown state and $z=\exp^{\frac{\mu}{\kappa T}}$ is the so--called fugacity.

Setting $U_{0}=0$ in Eq.\,(\ref{NPINT}), we are able to calculate the condensation temperature $T_{0}$ in the ideal case. To do this, we assume that the number of particles in the ground state in negligible, together with $\mu=0$ at the condensation temperature.
Thus, we obtain
\begin{equation}
\label{T0}
\kappa T_{0}=\hbar \omega_{0}\Bigg[\frac{2^{\frac{3}{2}} \Gamma(\frac{3}{2}+1)N}{2^{\frac{3}{s}} \Gamma(\frac{3}{s}+1)\zeta\Bigl({\frac{3}{s}+\frac{3}{2} }\Bigr)}\Bigg]^{\frac{1}{\frac{3}{s}+\frac{3}{2} }}
\end{equation} 
where we have used that $g_{\nu}(z=1)=\zeta(\nu)$, being $\zeta(\nu)$ the Riemann Zeta Function.

Since the total number of particles is given by $N = N_{0} + N_{e}$, where $N_{0}$ is the number of particles in the ground state and $N_{e}$ is the number of particles in the exited states, the condensate fraction can be written as $N_{0}/N = 1 - N_{e}/N$, and according to  (\ref{NPINT}) and (\ref{T0}), at temperatures $T$ below $T_{0}$ , the condensate fraction is given by:
\begin{equation}
\label{eq:CF}
\frac{N_{0}}{N}=1-\Bigg(\frac{T}{T_{0}}\Bigg)^{\frac{3}{2}+\frac{3}{s}},
\end{equation}
where as usual, we have used the fact that the number of particles that can be located in the excited states has a maximum
\begin{equation}
\label{eq:Ne_max}
N_{e}^{max}= \Big(\frac{\kappa T}{\hbar \omega_{0}}\Bigr)^{\frac{3}{2}+\frac{3}{s}} \frac{2^{\frac{3}{s}} \Gamma(\frac{3}{s}+1)}{2^{\frac{3}{2}} \Gamma(\frac{3}{2}+1)} \zeta \Bigl(\frac{3}{2}+\frac{3}{s}\Bigr)
\end{equation}
\begin{figure*}
\centering
\includegraphics[width=11.7cm]{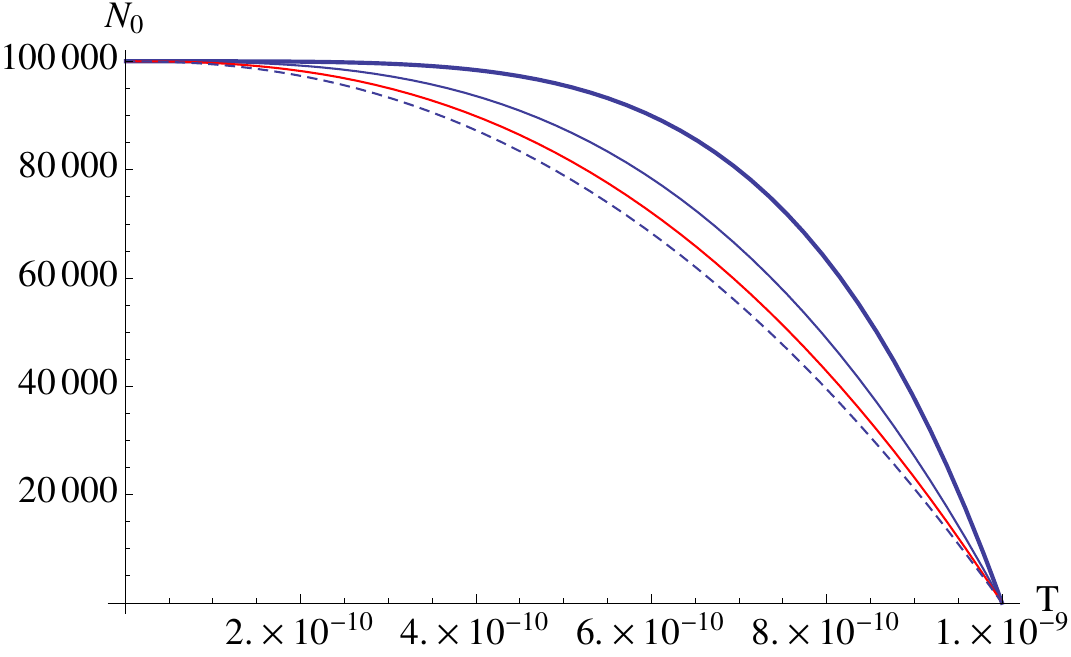}
\centering
\caption{\label{fig:N0} Number of particles in the condensed state as a function of the temperature for $a>0$, i. e., the interacting repulsive case, at temperatures smaller than the condensation temperature $T_{c}\sim 10^{-9}$\,K. The plot shows different values for the exponent $s$. The case for $s=1$ is represented by the solid blue line, the case $s=2$ by thin blue line, the case $s=3$ by the red line and the case $s=4$ by the dashed blue line. We observe that for a given temperature, the number of particles in the condensed state decreases as the exponent $s$ grows.}
\end{figure*}
In order to obtain the shift in the condensation temperature caused by interactions, let us expand expression (\ref{NPINT}) to first
order in $T=T_{0}$, $\mu=0$ and, $U_{0}=0$. Recalling that $T_{0}$ is the condensation temperature
given by expression (\ref{T0}) (see for instance Refs.\,\cite{eli,sal}), with the result :
\begin{equation}
\label{eq:CTI}
\frac{T_{c} -T_{0}}{T_{0}} =\frac{\Delta T_{c}}{T_{0}}=-\Bigg(\frac{a}{\sqrt{\hbar/m \omega_{0}}}\Bigg)\Lambda_{s} N^{\frac{s}{3(s+2)}},
\end{equation}
where
\begin{eqnarray}
\label{eq:CTI1}
\Lambda_{s} &=&\Bigg[\frac{2^{3/2}\Bigl(\zeta(\frac{3}{s}+\frac{1}{2})\zeta(\frac{3}{2})-G_{3/2,1/2,3/s}(1)\Bigr)}{\pi^{1/2}\zeta (\frac{3}{s}+\frac{3}{2})(\frac{3}{s}+\frac{3}{2})}\Bigg]\\\nonumber &\times&\Bigg[\frac{2^{3/2}\Gamma\Bigl(\frac{7}{2}\Bigr)}{2^{3/s}\Gamma\Bigl(\frac{3}{s}+1\Bigr)\zeta\Bigl(\frac{3}{s}+\frac{3}{2}\Bigr)} \Bigg]^{\frac{s}{3(s+2)}}.
\end{eqnarray}
Where  we have setting in Eq.\,(\ref{eq:CTI}) the condensation conditions in the case of weakly interacting systems, i.e., at the condensation temperature $T_{c}$ for large $N$, in the mean field approach the chemical potential takes the value  \cite{Dalfovo,Pethick}
\begin{equation}
\label{eq:muc}
\mu_{c}=2\,U_{0}n_{0}(r=0).
\end{equation}
 In other words, $n_{0}(r=0)=\lambda_{c}^{-3}\zeta(3/2)$ in
the large $N$ limit, which means that the critical density \emph{near} the center of the spherical power law potential for any $s$ is
the same as that of the uniform model. 
For instance, in the case $s=2$, i. e., the usual harmonic oscillator potencial the current high precision experiments for $^{39}_{19}K$, show that the shift in the condensation temperature with respect to the ideal result, caused by the interactions among the constituents of the gas is about $5\times10^{-2}$ with a 1\% of error \cite{RP}. In other words, when repulsive interactions ($a>0$) are present, the condensation temperature (depending of the exponent\,$s$) is smaller compared with the ideal case, see Eq.\,(\ref{eq:CTI}).

Additionally, when interactions are present the condensate fraction can be obtained from eqns. (\ref{NPINT}), as \\$N_{0}/N = 1 - N_{e}/N$, where, as in the non--interacting case the number of particles that can be located in the excited states can be expressed as
\begin{eqnarray}
\label{eq:NeI}
N_{e}^{max}&=&\Bigl(\frac{\kappa T}{\hbar \omega_{0}}\Bigr)^{\frac{3}{2}+\frac{3}{s}} \frac{2^{\frac{3}{s}} \Gamma(\frac{3}{s}+1)}{2^{\frac{3}{2}} \Gamma(\frac{3}{2}+1)}  \zeta \Bigl(\frac{3}{2}+\frac{3}{s}\Bigr) \nonumber\\ &\times& \Bigg[1-2U_{0}(\kappa T)^{\frac{1}{2}}\Bigl(\frac{m}{2\pi\hbar^{2}}\Bigr)^{\frac{3}{2}}\,\frac{G_{3/2,1/2,3/s}(1)}{ \zeta \Bigl(\frac{3}{2}+\frac{3}{s}\Bigr)}\Bigg], \,\,\,\, 
\end{eqnarray}

We show in Fig. \ref{fig:N0} the plot for some specific values of the exponent $s$. Moreover, we have used typical laboratory conditions, i. e., $N \sim 10^{5}$ particles, $a \sim10^{-9}$ m, and $m\sim 10^{-26}$ kg \cite{Dalfovo}, with a typical condensation temperature of order $T_{c}\sim 10^{-9}$ $K$. 

Note that when interactions are present and for temperatures smaller than $T_{c}$ the condensed fraction is also smaller compared with the ideal case for $a>0$. From this analysis we infer that when $s$ grows, the number of particles in the condensate decreases, which is in agreement with the results given by the dynamical analysis, in which we interpreted the increasing of $s$ as a increase of instability. For instance, the analysis performed in Sec.\,(\ref{sec:dynamical}) together with the results exposed in the present section, allow us to conclude that the cases $s=3$ and $s=4$ are less stable than the usual harmonic trap $s=2$. In this line of thinking, the case $s \rightarrow \infty$ which describes a box potential, is clearly, less stable than the aforementioned scenarios. Also, the spatial extend of the condensate deduced from the Thomas--Fermi approach decreases as the exponent $s$ grows. 


\section{Discussion}
\label{sec:Conclusions}

In order to study the stability of the condensate when the exponent of the trap varies, we have analyzed 
the stability conditions of a Bose--Einstein condensate in a spherical power law potential using techniques from dynamical systems. Also, we have confronted these results with the analysis of the corresponding condensed fraction, showing that both predictions are compatible. 

We observe that when the exponent $s$ grows, the system tends to be unstable, i.e., when the exponent 
fulfill the condition $s>2$. Basically, the perturbations deduced in the above sections are directly related to perturbations of the order parameter (see the system (\ref{per})), and consequently, related to fluctuations of the condensed state. This fact can be interpreted as follows: when $s$ grows, the number of particles in the ground state decreases and clearly the number of particles in the excited states increase. Also, the size of the cloud decreases as the exponent $s$ increases. Let us add, that according to our results the most stable scenarios for the condensate when a positive self--interaction parameter is assumed, are given for small enough values of the exponent $s$.

Let us mention here that according to our interpretation, i.e., that the instability means a loss of particles of the condensate when the exponent $s$ grows, does not means that the equilibrium condensate density in the instable cases, i.e. $s>2$ is zero, or at least does not allows to conclude this fact. As was analyzed in the previous sections, the loss of particles of the condensed state grows when $s$ grows. In other words, the so--called \emph{depletion} increases with the exponent $s$. A more suitable interpretation is that  the systems with bigger exponent $s$ are less stable than the systems with small exponent $s$. However, this fact do not implies that there are not a well defined equilibrium condensate densities for $s>2$. Moreover, as was mentioned previously in the manuscript, the limitations associated with the Thomas--Fermi approach used in the above analysis, allows us to deduce instabilities in a certain region, i.e., $\bar{r}<(2\bar{\mu})^{1/s}$. Thus, as a consequence we can not extrapolate completely our results to the entire condensate but can be used as an insight of the condensate stability.

In addition, the dynamical system formalism can extended to more general traps, even time--dependent traps and rotating traps in order to analyze the stability of the system under more general conditions, where it would be interesting to show if the solutions are independent of the initial conditions. Also, a numerical simulation can be used in order to match the results obtained in the present report with those eventually obtained in a numerical analysis. Clearly these topics deserve deeper investigation, which will be presented elsewhere.


%

\section{Acknowledgments}
E. C. and C. E-R acknowledges MCTP/UNACH for financial support. We thanks to P. Sloane for his opinion on the manuscript.

\section{Authors contributions}
All the authors were involved in the preparation of the manuscript and all authors contributed equally to the paper.
All the authors have read and approved the final manuscript.

\appendix 
\section{Dynamical system background}\label{app1}
The stability of a particular system of equations system can be studied through its critical points and performing linear perturbations
around the critical values of the form $\bold{x}=\bold{x_0}+\bold{u}$, where
$\bold{x}=(x,y,z)$ and $\bold{u}=(\delta x, \delta y, \delta z)$. The equations of motion (\ref{eq:systemdynamical}) can be written as $\bold{x}^{\prime} = \bold{f}(\bold{x})$, which upon linearization reads
\begin{equation}
\bold{u}^\prime = \mathcal{M} \bold{u}, \quad \mathcal{M}_{ij} =  \left. \frac{\partial f_{i}}{\partial x_{j}}\right|_{\mathbf{x}_{*}}, \label{per}
\end{equation}
where $\mathcal{M}$ is called the linearization matrix. The eigenvalues $m$ of $\mathcal{M}$ determine the stability of the critical points, whereas the eigenvectors determine the principal directions of the perturbations. In general, if $Re(m)<0$ \\ ($Re(m)>0$) the critical point is called stable (unstable). 
We should study the perturbations of the three dynamical variables $(x,y,z)$ in the phase space, but remember that they are not all independent because they are bound  by the Thomas--Fermi approximation (\ref{eq:TFA}), and the same happens to their perturbations. This constraint defines a two dimensional surface (\ref{eq:TFA_constraint}) upon which lie all physically relevant phase space trajectories (solutions), and then we will be interested on perturbations lying also on the constraint surface. In other words, perturbations which are perpendicular to the Thomas--Fermi constraint surface should be excluded from the analysis.

To resolve Eq.(\ref{per}) we list the eigenvalues of the stability matrix $\mathcal{M}$ for each of the critical point with the following ansatz at first order in the perturbations:
\begin{equation}
x=x_c +u, \quad y=y_c +v, \quad z=z_c +\omega,
\end{equation}
where
\begin{equation}
\xi^{\prime} = \mathcal{M} \xi, \quad \rightarrow \quad 
 \left(
    \begin {array}{c} 
u^\prime
\\\noalign{\medskip} v^\prime
\\\noalign{\medskip}\omega^\prime
\end {array} \right)  \, = \mathcal{M}  \left(
    \begin {array}{c} 
u
\\\noalign{\medskip} v
\\\noalign{\medskip}\omega
\end {array} \right)  \,
\end{equation}
is the system to resolve which has solutions
\begin{subequations}
\begin{eqnarray}
u&=& u_{+} e^{m_{+}\bar{r}} +u_{-} e^{m_{-}\bar{r}}, \\ \label{eq:solutions_dynamical-a}
v&=& v_{+} e^{m_{+}\bar{r}} +v_{-} e^{m_{-}\bar{r}}, \\ \label{eq:solutions_dynamical-b}
\omega&=& \omega_{+} e^{m_{+}\bar{r}} +\omega_{-} e^{m_{-}\bar{r}},  \label{eq:solutions_dynamical-c}
\end{eqnarray} \label{eq:solutions_dynamical}
\end{subequations}
where $[m_{+},m{-}]$ are the eigenvalues of the system and $\xi$ the eigenvectors and $[u_{+}, u_{-}, v{+}, v{-}, w_{+}, w_{-}]$ are arbitrary constants which are determined by the initial conditions.

\end{document}